\documentclass[aps, prl, preprint, superscriptaddress, a4paper, 11pt, floatfix]{revtex4-2}

\usepackage[LGRgreek]{mathastext}

\usepackage{graphicx}
\usepackage{epstopdf}
\usepackage{amsmath}
\usepackage{silence}

\usepackage[usenames,dvipsnames]{color}
\usepackage[hidelinks,colorlinks,linkcolor=blue,citecolor=blue,urlcolor=blue]{hyperref}

\usepackage{float}
\usepackage{geometry}
\usepackage{xspace}

\newcommand{\nb }{$2H$-NbSe$_2$\xspace}

\begin{document}

\title{Probing charge order of monolayer NbSe$_2$ within a bulk crystal}

\author{Doron Azoury}
\altaffiliation{These authors contributed equally to this work.}
\affiliation{Department of Physics, Massachusetts Institute of Technology, Cambridge, Massachusetts, 02139 USA}

\author{Edoardo Baldini}
\altaffiliation{These authors contributed equally to this work.}
\affiliation{Department of Physics, Massachusetts Institute of Technology, Cambridge, Massachusetts, 02139 USA}

\author{Aravind Devarakonda}
\affiliation{Department of Physics, Massachusetts Institute of Technology, Cambridge, Massachusetts, 02139 USA}

\author{Jiarui Li}
\affiliation{Department of Physics, Massachusetts Institute of Technology, Cambridge, Massachusetts, 02139 USA}

\author{Shiang Fang}
\affiliation{Department of Physics, Massachusetts Institute of Technology, Cambridge, Massachusetts, 02139 USA}

\author{Pheona Williams}
\affiliation{Department of Physics, Massachusetts Institute of Technology, Cambridge, Massachusetts, 02139 USA}

\author{Riccardo Comin}
\affiliation{Department of Physics, Massachusetts Institute of Technology, Cambridge, Massachusetts, 02139 USA}

\author{Joseph Checkelsky}
\affiliation{Department of Physics, Massachusetts Institute of Technology, Cambridge, Massachusetts, 02139 USA}

\author{Nuh Gedik}
\affiliation{Department of Physics, Massachusetts Institute of Technology, Cambridge, Massachusetts, 02139 USA}

\date{\today}

  \begin{abstract}
\textbf{
Atomically thin transition metal dichalcogenides can exhibit markedly different electronic properties compared to their bulk counterparts. In the case of \nb, the question of whether its charge density wave (CDW) phase is enhanced in the monolayer limit has been the subject of intense debate, primarily due to the difficulty of decoupling this order from its environment. Here, we address this challenge by using a misfit crystal that comprises NbSe$_2$ monolayers separated by SnSe rock-salt spacers, a structure that allows us to investigate a monolayer crystal embedded in a bulk matrix. We establish an effective monolayer electronic behavior of the misfit crystal by studying its transport properties and visualizing its electronic structure by angle-resolved photoemission measurements. We then investigate the emergence of the CDW by tracking the temperature dependence of its collective modes. Our findings reveal a nearly sixfold enhancement in the CDW transition temperature, providing compelling evidence for the profound impact of dimensionality on charge order formation in NbSe$_2$.
  }

\end{abstract}

\maketitle

Over the past decade, significant efforts have been made to explore the influence of reduced dimensionality on the emergence of collective phenomena, most notably in transition metal dichalcogenides (TMDs) hosting CDW and superconducting phases~\cite{sajadi2018gate,yu2019high,xi2016ising,feng2018electronic,sugawara2016unconventional,lu2015evidence}. A prototypical example is that of \nb. In the bulk form, this crystal exhibits a nearly commensurate 3 $\times$ 3 CDW phase ($T_{CDW}$ = 33 K) and a low-temperature superconducting order ($T_C$ = 7.2 K). There is a wide consensus that superconductivity is suppressed when this material is thinned down to the monolayer limit, emerging at transition temperatures on the order of 3 K. However, the fate of the CDW order in a two-dimensional crystalline environment remains unclear~\cite{xi2015strongly,ugeda2016characterization,lin2020patterns,xie2021charge,xi2016gate,lian2018unveiling,bianco2020weak,calandra2009effect,lin2022axial,chen2020visualizing}. Scanning tunneling microscopy experiments performed on monolayer \nb have shown a slight reduction in T$_{CDW}$ when the material is grown via molecular beam epitaxy on bilayer graphene \cite{ugeda2016characterization}. In contrast, Raman measurements of the exfoliated material on top of a sapphire substrate have reported a significant enhancement of $T_{CDW}$ up to 145 K~\cite{xi2015strongly}. Despite subsequent experimental and theoretical work attempting to resolve the controversy, the question of whether the CDW of \nb is enhanced or suppressed at the monolayer limit remains unanswered. The monolayer quality and durability, as well as substrate-related factors such as charge transfer, doping, screening, and strain, are believed to play a central role in these contradictory observations. These factors are challenging to prevent or minimize in a single monolayer on a substrate, adding significant difficulty to elucidating the intrinsic CDW properties of this system.

\begin{figure}
\begin{center}
  \centering{\includegraphics*[width=0.7\columnwidth]{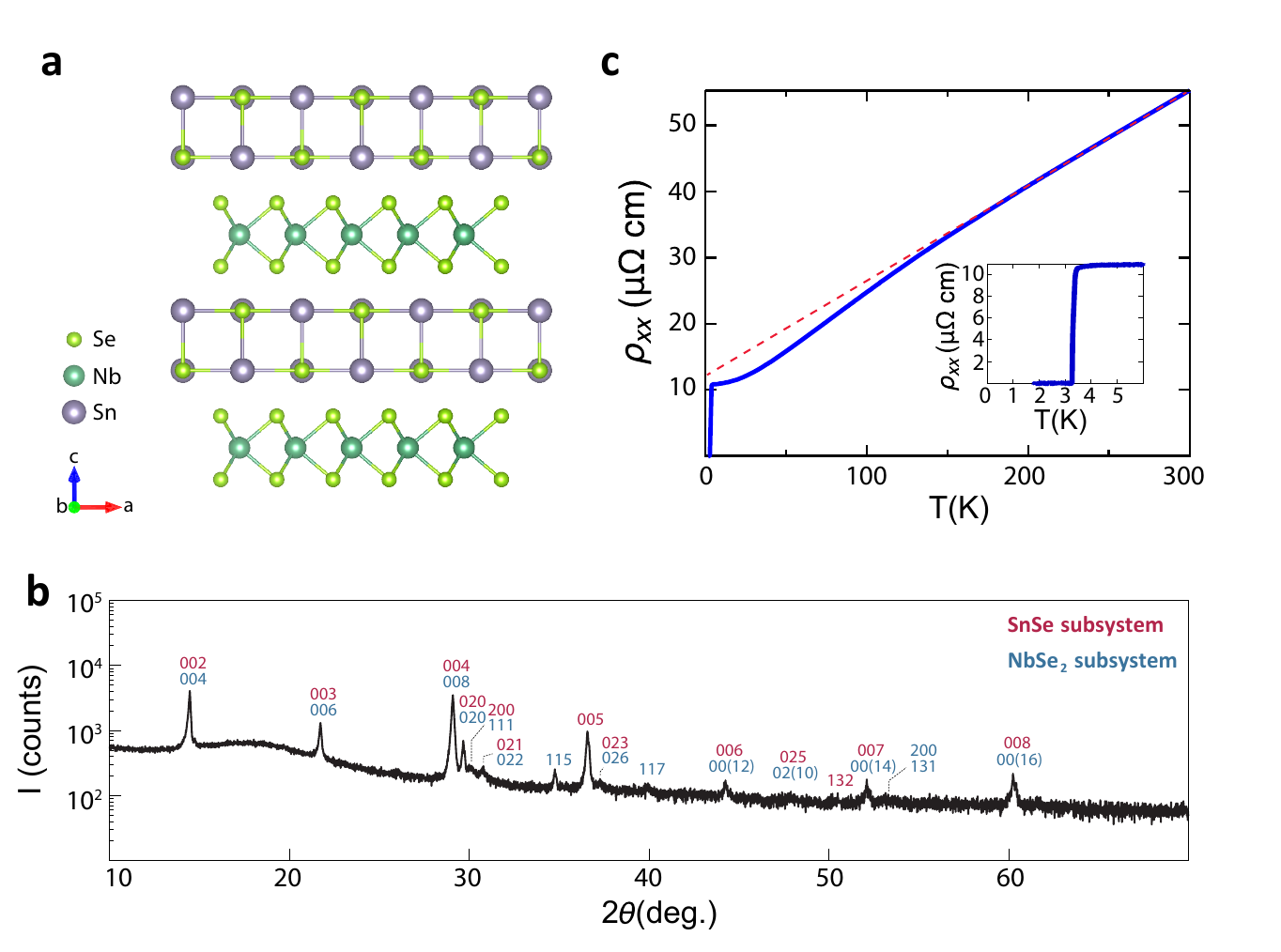}}
  \caption{{\bf Crystal structure and resistivity of [SnSe]$_{1.16}$NbSe$_2$.} {\bf a}, Diagram (side view) of the crystal structure of [SnSe]$_{1.16}$NbSe$_2$. {\bf b}, Powder x-ray diffraction of [SnSe]$_{1.16}$NbSe$_2$, the peaks that relate to the SnSe and NbSe$_2$ subsystems are indicated. {\bf c}, Resistivity as a function of temperature of [SnSe]$_{1.16}$NbSe$_2$, showing superconductivity transition temperature of 3.2K. The dashed red line is a guide to the eye, representing the linear response at the high temperature range.   }
  \label{fig1}
\end{center}
\end{figure}

Here, we investigate the impact of dimensionality on the CDW in \nb by utilizing a misfit crystal of [SnSe]$_{1.16}$NbSe$_2$~\cite{ng2022misfit}. This incommensurate structure consists of single-layer \nb separated by rock-salt spacers of SnSe (see Fig.~\ref{fig1}a and Supplementary Information). We demonstrate that such superstructure enables us to effectively study the intrinsic properties of monolayer \nb. To achieve this, we first identify the material's superconducting critical temperature (T$c$) through transport measurements and measure its low-energy band structure using angle-resolved photoemission spectroscopy (ARPES). Finally, we probe the emergence of the CDW by tracking its collective mode dynamics as a function of temperature via spontaneous Raman scattering. Our findings reveal that T$_{CDW}$ is considerably enhanced in comparison to bulk \nb, indicating that reduced dimensionality has a significant impact on the CDW phase. These results, obtained in a protected monolayer within the misfit bulk crystal, offer a clean approach to studying the CDW in the monolayer limit while minimizing environmental effects.

\begin{figure*}
\begin{center}
  \centering{\includegraphics*[width=1\columnwidth]{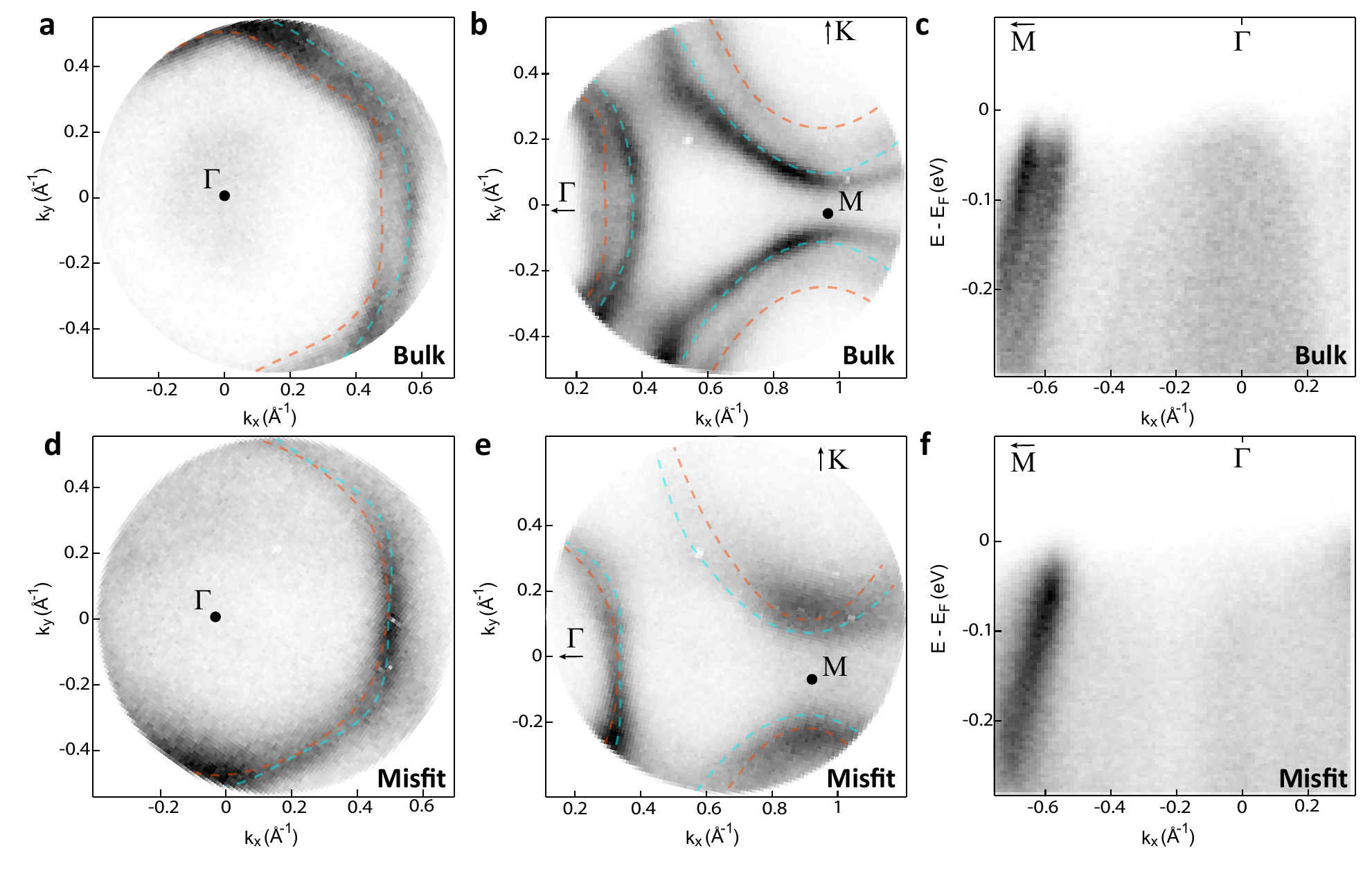}}
  \caption{{\bf Electronic band-structure of [SnSe]$_{1.16}$NbSe$_2$.} {\bf a-f}, ARPES measurements of bulk \nb and misfit crystal [SnSe]$_{1.16}$NbSe$_2$. Constant energy cuts at the Fermi level ({\bf a} and {\bf b} for the bulk, {\bf d} and {\bf e} for the misfit crystal). The orange cyan dashed lines corresponds to a DFT calculation (see Supplementary Information). Energy-momentum cut along the $\Gamma$-M high symmetry line ({\bf c} for the bulk and {\bf f} for the misfit crystal). }
  \label{fig2}
\end{center}
\end{figure*}

We first evaluate the conduction properties and electronic structure of the misfit crystal to demonstrate its effective two-dimensional behavior. Figure~\ref{fig1}c shows the material's resistivity as a function of temperature (see Supplementary Information). We find that T$_C$ is $\sim$3.2~K, consistent with values reported in monolayer \nb systems~\cite{xi2015strongly,xi2016ising} and [SnSe]$_{1.16}$NbSe$_2$~\cite{bai2018superconductivity}.

Next, we compare the electronic band structure of the misfit crystal with those expected for bulk and monolayer \nb. To this end, we performed ARPES measurements of both [SnSe]$_{1.16}$NbSe$_2$ and bulk \nb under the same experimental conditions (see Supplementary Information). In Figs.~\ref{fig2}a-c, we present the results for the bulk crystal, where we observe two hexagonal contours around the $\Gamma$ point and two warped triangular contours around the K point, all of which are associated with Nb 4d bands. Additionally, the signal at the zone center corresponds to Se 4p$_z$ derived 'pancake' surface. These observations are consistent with previous measurements of bulk \nb ~\cite{inosov2008fermi,rossnagel2001fermi}. Examination of the measured band structure of the misfit crystal (Figure~\ref{fig2}d-e) reveals three features that indicate behavior alike that of monolayer \nb. Firstly, the misfit lacks a large inter-layer driven splitting of the contours around the $\Gamma$ point, which instead appears for the bulk. Secondly, we observe a circular inner contour around the $\Gamma$ point rather than the hexagonal one seen in the bulk. Thirdly, the spectral weight at the zone center is absent, which is consistent with the fact that the Se 4p$_z$ band does not cross the Fermi level in the monolayer limit. These observations are in excellent agreement with density functional theory calculations for a bulk and monolayer \nb, which are superimposed on the experimental data in Fig.~\ref{fig2}. In general, charge transfer from the rock-salt layer may dope the TMD layer, as previously demonstrated in other misfit compounds \cite{leriche2021misfit}. The comparison with the calculated Fermi surface indicates negligible charge transfer in our case, which is also consistent with previous studies \cite{wiegers1991misfit,wiegers1995charge}, making it a faithful representation of an isolated monolayer \nb. It is worth noting that the energy resolution of our ARPES measurements ($\Delta$E $\approx$ 60 meV) did not allow us to detect CDW-induced changes in the band structure.

Both the measured conductivity and electronic band structure establish the presence of an effective monolayer \nb within the misfit bulk crystal. This allows us to probe the CDW in \nb at the monolayer limit while the monolayer resides within a bulk crystal environment. 


An electron-phonon-mediated CDW phase can be probed by monitoring its collective excitations -- the amplitude and phase modes -- typically via Raman or infrared spectroscopy, respectively. Here, we perform back to back Raman measurements of [SnSe]$_{1.16}$NbSe$_2$ and bulk \nb and observe the CDW amplitude mode in both crystals as a function of temperature (see experimental details in Supplementary Information). Figure~\ref{fig3} shows the temperature dependent Raman susceptibility in the bulk and misfit crystals, taken with crossed polarization configuration.

\begin{figure*}
\includegraphics*[width=0.95\columnwidth]{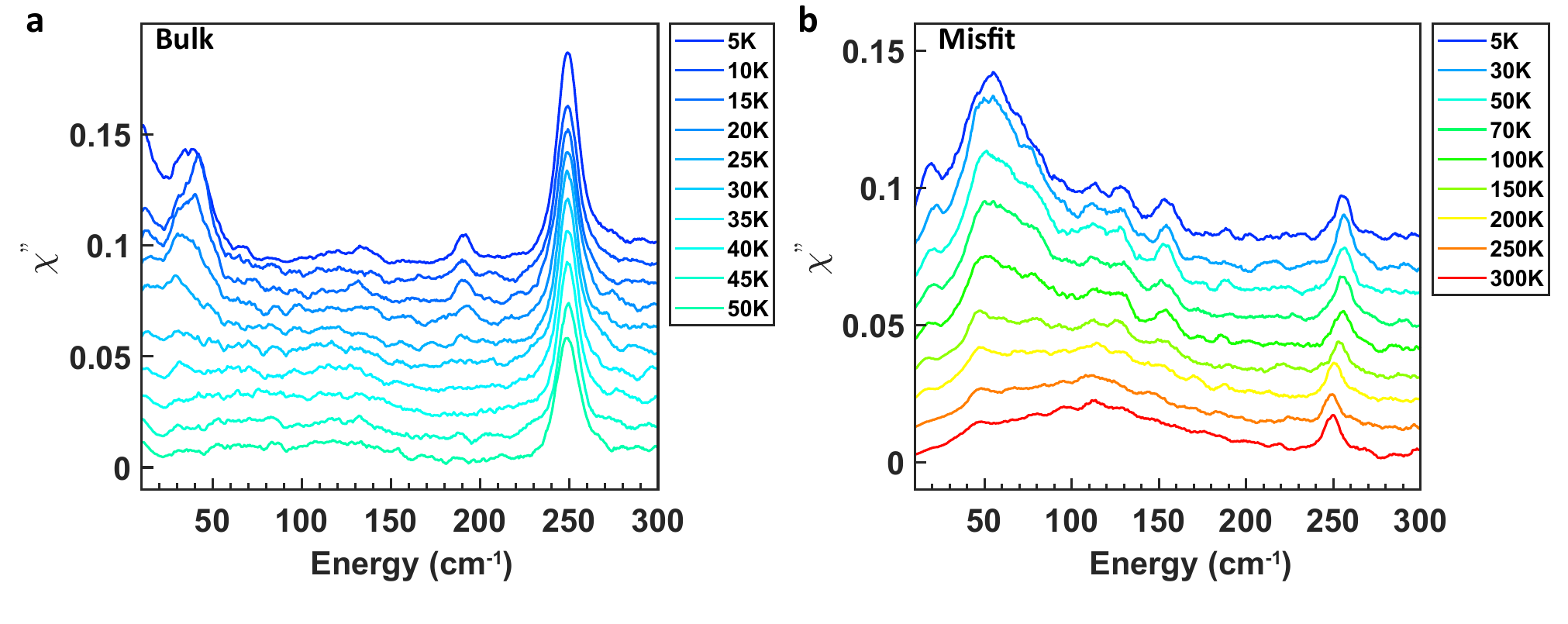}
\caption{{\bf Raman measurements.} {\bf a}, Temperature-dependent Raman measurements of bulk \nb, performed in crossed polarization configuration. A temperature-independent E$_{2g}$ phonon mode appears around 250~cm$^{-1}$, and a softening CDW amplitude mode appears below approximately 30K. {\bf b}, Same measurement as in {\bf a}, but for the misfit crystal [SnSe]$_{1.16}$NbSe$_2$. As in case of bulk \nb, a temperature-independent E$_{2g}$ phonon mode appears at 250~cm$^{-1}$, however an amplitude mode becomes visible at a much higher temperature of about 250K. In both {\bf a} and {\bf b} the temperature dependent spectra are shifted vertically for clarity.  }
\label{fig3}
\end{figure*}

At room temperature, the Raman spectrum of the misfit crystal resembles the bulk Raman spectrum at T$>$T$_{CDW}$, with the appearance of the E$_{2g}$ phonon mode around 250~cm$^{-1}$. Upon cooling below the bulk CDW transition temperature of 33.5~K, we observe the expected Raman response of bulk \nb ~\cite{sooryakumar1980raman,measson2014amplitude}: it develops a strong CDW amplitude mode centered around 40~cm$^{-1}$, which softens in the temperature range of 30~K and 5~K, and an additional temperature dependent peak at 190~cm$^{-1}$. On the other hand, the misfit crystal develops a strong new Raman mode centered around 50~cm$^{-1}$ at a significantly higher temperature of about 250~K. Besides the modes that are associated with the \nb layer, we identify 3 phonon modes of SnSe~\cite{pal2020pressure} at 109~cm$^{-1}$, 131~cm$^{-1}$ and 153~cm$^{-1}$. 

\begin{figure*}
\begin{center}
  \centering{\includegraphics*[width=0.95\columnwidth]{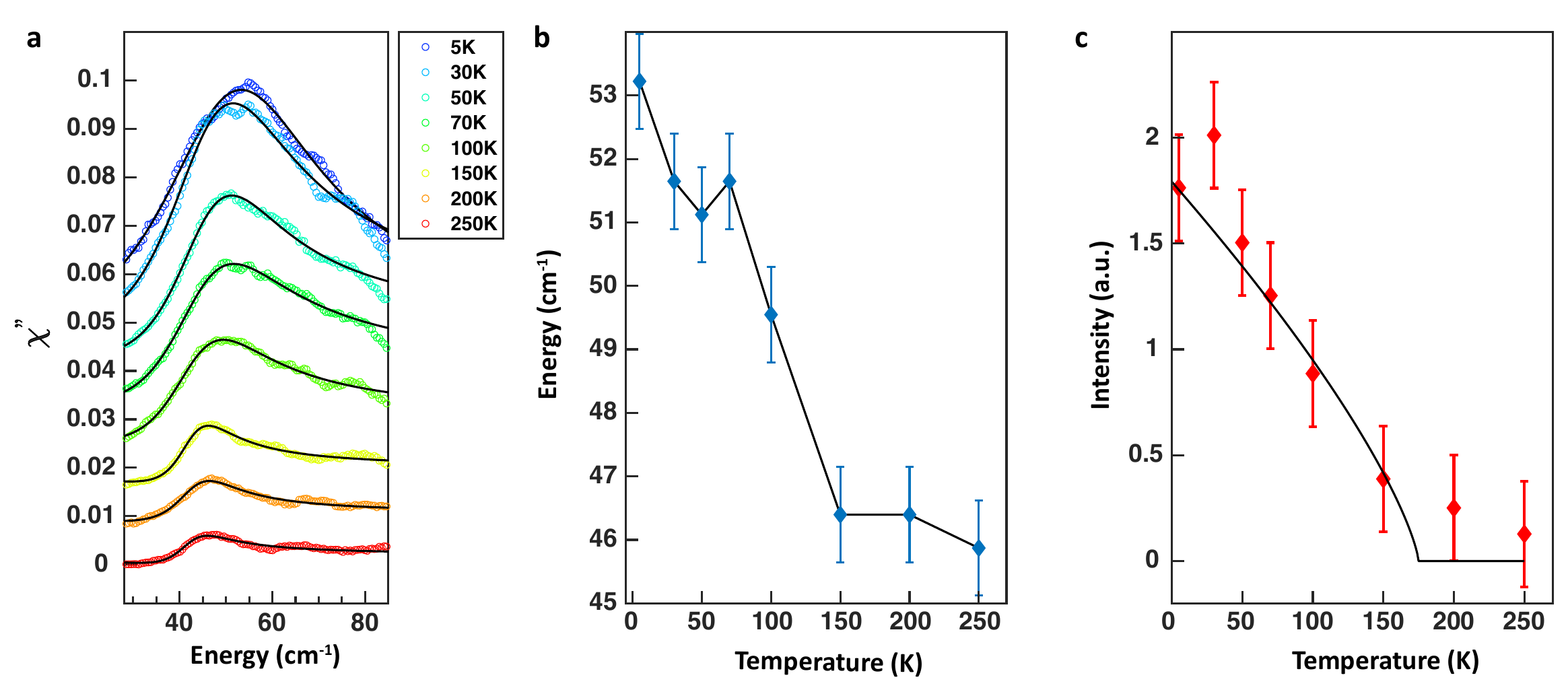}}
  \caption{{\bf Amplitude mode in [SnSe]$_{1.16}$NbSe$_2$.} {\bf a}, A close-up on the energy range of the amplitude mode as a function of temperature in the misfit crystal. The black line is a Fano line-shape fit, performed after temperature dependent background subtraction. {\bf b}, Central energy of the amplitude mode as a function of temperature. {\bf c}, Intensity of the amplitude mode as a function of temperature. The black line is a mean-field calculation.  }
  \label{fig4}
\end{center}
\end{figure*}

Figure~\ref{fig4}a shows the temperature dependence of the amplitude mode after background subtraction. A fit of the mode (black lines in Figure~\ref{fig4}a) shows a clear softening within the temperature range of 250~K and 5~K, where the temperature dependent central energy is presented in Figure~\ref{fig4}b. Such softening indicates a shallower free energy potential surface for elevated temperatures, and identifies this peak with the amplitude mode of the CDW. In order to identify T$_{CDW}$, we plot in Figure~\ref{fig4}c the amplitude mode intensity as a function of temperature. We find good agreement with a mean field calculation (black line in Figure~\ref{fig4}c, see Supplementary Information), which allows us to determine a transition temperature of about 175~K. An indirect indication for the enhanced CDW can be found by a closer examination of the resistivity measurements, manifested as a deviation from a linear response (represented by the red dashed line in Fig. \ref{fig1}c) starting at approximately 150 K. Such a kink is reminiscent, for example, of the transport signatures of the pseudogap in cuprates which as in the case of \nb becomes partially gaped below the transition temperature. 

Indeed, as mentioned earlier, previous Raman measurements of exfoliated monolayer \nb showed a very similar temperature-dependent amplitude mode~\cite{xi2015strongly,lin2020patterns}, indicating a dramatic enhancement of the CDW phase. However, since the observed CDW enhancement in our experiments is based on a monolayer within a bulk crystal, we can safely rule out environmental factors that may induce such enhancement. This in turn leads us to determine that the origin of the enhancement is the reduced dimensionality of \nb. Our finding resolves the main debate point between the contradicting reported results\cite{ugeda2016characterization,xi2015strongly,lian2018unveiling,bianco2020weak} for the CDW properties at the monolayer limit.  

The CDW phase forms once the system can find a lower energy ground state by  mutual distortion of the charge distribution and the lattice~\cite{gruner1988dynamics}. This condition is realized when the electronic energy gain is greater than the lattice elastic energy loss. In the case of bulk \nb, the inter-layer coupling contributes to the lattice elastic energy, meaning it makes it harder for the lattice to deform. In the monolayer limit, the inter-layer coupling does not play a role, which may increase the CDW stability~\cite{lian2018unveiling}. Examination of the crystal structure of [SnSe]$_{1.16}$NbSe$_2$~\cite{ng2022misfit} shows a significantly larger separation of single layers \nb compared to the van der Waals spacing of bulk \nb by a factor of $\sim$ 3. Such a large separation practically decouples the single layers, and therefore reduces the elastic energy cost of the CDW phase transition. As seen in previous studies~\cite{xi2015strongly,lin2020patterns}, this leads to a significant enhancement of the CDW in an exfoliated monolayer \nb, whereas in the current study we show that the enhancement is a direct consequence of reduced dimensionality by probing the CDW at the monolayer limit within a protected misfit crystal.   

\clearpage
\subsection*{Data availability}
The data that support the plots within this paper and other finding of this study are available from the corresponding author upon reasonable request.

\subsection*{Acknowledgements}

The authors thank Dongsung Choi for technical assistance during the ARPES measurements. D.A., E.B. and N.G. acknowledge support by the US Department of Energy, BES DMSE (data taking, analysis and manuscript writing) and Gordon and Betty Moore Foundation’s EPiQS Initiative grant GBMF9459 (instrumentation). D. A. acknowledges financial support by the Zuckerman STEM Leadership Program. P.W., A.D. and J.C. acknowledge support, in part, by the Gordon and Betty Moore Foundation EPiQS Initiative, Grant No. GBMF9070 to J.G.C (instrumentation development, DFT calculations), the US Department of Energy (DOE) Office of Science, Basic Energy Sciences, under award DE-SC0022028 (material development), and the Office of Naval Research (ONR) under award N00014-21-1-2591 (advanced characterization). P.O.W. acknowledges support from the STC Center for Integrated Quantum Materials (NSF grant DMR-1231319). J.L. and R.C. acknowledge support by the National Science Foundation under Grant No. 1751739.

\subsection*{Contributions}

D.A. and E.B. performed the ARPES measurements; A.D. and P.W. synthesized the crystals and performed the transport measurements; J.L. performed the Raman measurements; S.F. performed the DFT calculations; D.A. analyzed the data and wrote the manuscript with contributions from all authors; R.C., J.C. and N.G. supervised the project.

\subsection*{Competing interests}
The authors declare no competing interests.

\subsection*{Corresponding author}
Nuh Gedik gedik@mit.edu

\bibliography{MisfitCDW}

\end{document}